\documentstyle[preprint, eqsecnum,aps]{revtex}
\tightenlines
\begin{document}
\draft
\title{Light-Cone Quark Model Analysis of 
       Radially Excited Pseudoscalar and Vector 
       Mesons}
\author{Daniel Arndt and Chueng-Ryong Ji}
\address{Department of Physics, North Carolina State University,
         Raleigh, N.C. 27695-8202}
\maketitle
\begin{abstract}
We present a relativistic constituent quark model to analyze the 
mass spectrum and hadronic 
properties of radially excited $u$ and $d$ quark sector mesons.
Using a simple Gaussian function as a trial wave 
function for the variational principle of
a QCD motivated Hamiltonian, we obtain the mass 
spectrum consistent with the experimental data. 
To do the same for 
several observables such as decay constants and 
form factors, it seems necessary to include both
Dirac and Pauli form factors on the level of constituent
quarks. 
Taking into account the quark form factors, 
we thus present the generalized formulas for the rho meson
decay constant and form factors as well as
the $\pi\gamma$ transition form factor.
We also
predict several hadronic properties for the radially
excited states. 
\end{abstract}

\pacs{12.39.Ki,13.40.Gp,13.40.Hq,14.40.-n}

\section{Introduction}
Based on 
a detailed analysis of the meson mass spectrum by Godfrey
and Isgur (GI)~\cite{GoI85},
the light-cone~(LC) approach has been adopted
to describe the pion 
decay constant~\cite{CGN94}, the charge form
factor of the pion~\cite{CGN94,CGN95b,CGN95c,CGN95d}, 
the electromagnetic form factors of the $\rho$ 
meson~\cite{CGN95a} and the radiative $\rho\pi$ and $\omega\pi$ 
transition form factors~\cite{CGN95b,CGN95c}. 
In this approach, 
Cardarelli {\it et~al.}\ used the eigenfunctions of the effective
$q\bar{q}$ GI Hamiltonian
as orbital wave functions which consist
of a truncated set of at least 38 harmonic oscillator~(HO)
basic states. 
While these
wave functions reproduce the meson mass spectrum very well,
it became necessary to introduce Dirac and Pauli 
form factors on the level of the constituent quarks~(CQ)
in order to reproduce hadronic meson properties. 
This has been done to 
calculate the charge form factor of the pion, 
the radiative $\rho\pi$ and $\omega\pi$ 
transition form factors~\cite{CGN94,CGN95c} (both including
Dirac and Pauli CQ form factors) and the electromagnetic 
form factors of the $\rho$ meson~\cite{CGN95a} (Dirac CQ
form factor only).

Recently~\cite{ChJ98}, however, it was observed that 
both the masses and the 
hadronic 
properties of ground state pseudoscalar and vector
mesons can fairly well be reproduced 
by taking just a single $1S$-state 
HO wave function 
given by
\begin{equation} \label{E:phi_1}
  \phi_{1S}(\vec{k}^2)
    =\frac{1}{\pi^{3/4}\beta^{3/2}}
     \exp\left(-\frac{k^2}{2\beta^2}\right) 
.\end{equation}
In these calculations, the choice of the Gaussian 
parameter $\beta$ has been
obtained by the variational principle for the
QCD motivated Hamiltonian including the Coulomb
plus confining potential.
In this paper, we adopt this model~\cite{ChJ98}
and extend it to the radially excited
states for the $u$ and $d$ quark sector
where
the masses of $u$ and $d$ CQ can be taken
equal, $m_u=m_d$.
The necessity of CQ Dirac and Pauli form factors
was also confirmed in our analysis of
both ground state and excited states.
We thus present the generalized formulas
for the rho meson decay constant and form factors
as well as the $\pi\gamma$ transition form factor
taking 
into account both Dirac and Pauli CQ form factors.
Following the ground state analysis,  
it may not be so unreasonable to take
$2S$ and $3S$ HO wave functions
given by
\begin{equation} \label{E:phi_2}
  \phi_{2S}(\vec{k}^2)
    =\frac{1}{\sqrt{6}\pi^{3/4}\beta^{7/2}}
     \left(-3\beta^2+2k^2\right)
     \exp\left(-\frac{k^2}{2\beta^2}\right)
\end{equation}
and
\begin{equation} \label{E:phi_3}
  \phi_{3S}(\vec{k}^2)
    =\frac{1}{2\sqrt{30}\pi^{3/4}\beta^{11/2}}
     \left(15\beta^4-20\beta^2k^2+4k^4\right)
     \exp\left(-\frac{k^2}{2\beta^2}\right) 
\end{equation}
for the first and second excited states, respectively.
However, we tried the variations from these wave functions
and obtained more 
justifications for using Eqs.~(\ref{E:phi_2}) and 
(\ref{E:phi_3}), as we will present in the
next section.

The LC formalism (for a recent review see~\cite{BPP97}),
first introduced by Dirac~\cite{Dir49},
presents a natural framework to include
the relativistic effects which turned out to be crucial to 
describe the low-lying mesons~\cite{GoI85}.
Distinctive features of the LC time ($\tau=t+z/c$)
quantization include the suppression of vacuum 
fluctuations~\cite{JiR96}
and the conversion of the dynamical problem from boost
to rotation~\cite{JiS92}, 
which has the compact parameter space. 
Various 
hadronic properties are calculated using the 
well established formulation in the Drell--Yan--West
$q^+=0$ frame~\cite{DrY70}.
We also use the 
Melosh transformation~\cite{Mel74} to assign proper
quantum numbers $J^{PC}$ to
the mesons.

In our model, the meson state $|M\rangle$ is 
represented by
\begin{equation}
  |M\rangle=\Psi_{Q\bar{Q}}^M|Q\bar{Q}\rangle
,\end{equation}
where $Q$ and $\bar{Q}$ are the effectively dressed quark and
antiquark. 
The model wave function consists of the radial wave function
$\phi_{nS}$, the spin-orbit wave function $\mathcal{R}$
obtained by the Melosh-transformation
and a Jacobi factor
\begin{equation} \label{E:Psi_QQ_M}
  \Psi_{Q\bar{Q}}^M=
  \Psi(x, \vec{k}_\perp, \lambda_q, \lambda_{\bar{q}})=
  \sqrt{\frac{\partial k_n}{\partial x}}
  {\mathcal R}(x, \vec{k}_\perp, \lambda_q, \lambda_{\bar{q}})
  \phi_{nS}(x,\vec{k}_\perp)
.\end{equation}  
Here, the Jacobian of the 
variable transformation 
$\{k_n,\vec{k}_\perp\}\to\{x,\vec{k}_\perp\}$ is given by
$\frac{\partial k_n}{\partial x}=\frac{M_0}{4x(1-x)}$
and the spin-orbit wave function for spin ($S$, $S_z$) is given by
\begin{eqnarray}
  {\mathcal R}_{S,S_z}(x, \vec{k}_\perp, 
                      \lambda_q,\lambda_{\bar{q}})
  =
   \sum_{\lambda\lambda'} &&
   \left<\lambda_q\left|
     {\mathcal R}^{\dagger}_M(x,\vec{k}_\perp, m_q)
   \right|\lambda\right>
   \left<\lambda_{\bar{q}}\left|
     {\mathcal R}^{\dagger}_M(1-x,-\vec{k}_\perp, m_{\bar{q}})
   \right|\lambda'\right>
         \nonumber \\
         &&
   \times
   \left<
     \frac{1}{2}\frac{1}{2},\lambda\lambda'|SS_z
   \right>
,\end{eqnarray} 
where the Melosh rotation is given by
\begin{equation}
  {\mathcal R}^{\dagger}_M(x,\vec{k}_\perp, m)
  =
  \frac{m+xM_0-i\sigma(\hat{n}\times\hat{k})}
       {\sqrt{(m+xM_0)^2+\vec{k}_\perp^2}} 
\end{equation}
with $\hat{n}=(0,0,1)$.
Also, the wave function is normalized as
\begin{equation}
  \sum_{\nu\bar{\nu}}\int d^3k
  \left|
    \Psi_{Q\bar{Q}}^M(x, \vec{k}_\perp, \nu, \bar{\nu})
  \right|^2
  =1  
,\end{equation}
where $\vec{k}=(k_n, \vec{k}_{\perp})$ and 
$k_n=(x-1/2)M_0$.

We treat the meson mass according to the invariant meson mass
scheme where the meson mass square $M_0^2$ for the case
$m_q=m_{\bar{q}}$ is given by
$M_0^2=\frac{\vec{k}_\perp^2+m_q^2}{x(1-x)}$
with $m_q$ being the CQ mass. 
The intrinsic LC variables $x$ and $\vec{k}_\perp$ are  
given by
$x=k_q^+/P^+=1-k_{\bar{q}}^+/P^+$
and
$\vec{k}_\perp=\vec{k}_{q\perp}-x\vec{P}_\perp
              =-\vec{k}_{\bar{q}\perp}+(1-x)\vec{P}_\perp$.
Here, the subscript $\perp$ indicates the component 
perpendicular
to the LC quantization axis $\hat{n}$ and the $+$~component of
a 4-vector $k=(k^0,\vec{k})$ is given by 
$k^+=k^0+\hat{n}\cdot\hat{k}$. The total 4-momentum of a meson
in the frame of $\vec{P}_\perp=\vec{0}_\perp$ 
is therefore given by 
$P=(P^+,M^2/P^+,\vec{0}_\perp)$ and those of the quark and
antiquark are given by 
$k_q=\left(xP^+,\frac{m_q^2+\vec{k}_\perp^2}{xP^+},
           \vec{k}_\perp\right)$
and
$k_{\bar{q}}=\left((1-x)P^+,
             \frac{m_{\bar{q}}^2+\vec{k}_\perp^2}{xP^+},
             -\vec{k}_\perp\right)$,
respectively.

The paper is organized as follows: 
In Sec.~\ref{S:fixing_model_parameters}, 
we fix our model parameters.
First, in Sec.~\ref{S:CQ_mass_and_potential_parameters},
we set up the QCD motivated
effective Hamiltonian for the $q\bar{q}$ interaction. Using
the wave function in Eq.~(\ref{E:phi_1}) we determine the Gaussian 
parameter $\beta$ from the variational principle. With a 
relativistic hyperfine interaction we then determine
the best potential parameters and CQ mass $m_q$ to 
fit the
mass spectrum for the radially excited states 
of the $u$ and $d$ quark sector
as well as the pion decay constant $f_\pi$.
In Sec.~\ref{S:CQ_form_factor_parameters}, 
we introduce the CQ form factors
and fix the parameters associated with
these form factors using the pion form 
factor $F_\pi$ and the transition magnetic moments
$\mu_{\rho\pi}$ and $\mu_{\omega\pi}$. 
In Sec.~\ref{S:hadronic_properties},
we present the generalized formulas taking into account
the CQ form factors and
predict the rho decay constant $f_\rho$,
the rho electromagnetic form factors
and the $\pi^0\to\gamma^*\gamma$ transition
form factor $F_{\pi\gamma}$.
We also predict 
several hadronic observables
involving radially excited states
in this section.
Summary and discussion follow in 
Sec.~\ref{S:summary}.
In the Appendix, 
the details of the fitting procedure 
for the CQ structure
parameters are presented.

\section{Fixing the Model Parameters} 
\label{S:fixing_model_parameters}

\subsection{CQ Mass and Potential Parameters} 
\label{S:CQ_mass_and_potential_parameters}
We start with the QCD motivated effective $q\bar{q}$ Hamiltonian
as given by~\cite{ChJ98,GoI85}
for the description of the
meson mass spectrum
\begin{equation}
  H_{q\bar{q}}\left|\Psi_{nlm}^{SS_z}\right>
    =\left(H_0+V_{q\bar{q}}\right)\left|\Psi_{nlm}^{SS_z}\right>
    =M_{q\bar{q}}\left|\Psi_{nlm}^{SS_z}\right>
,\end{equation}
where $M_{q\bar{q}}$ is the mass of the meson, 
the free Hamiltonian is
$H_0=\sqrt{m_q^2+\vec{k}^2}+\sqrt{m_{\bar{q}}^2+\vec{k}^2}$,  
$|\Psi_{nlm}^{SS_z}>$ is the meson wave function as given 
in Eq.~(\ref{E:Psi_QQ_M}) 
and $V_{q\bar{q}}$ is the addition of the central 
potential $V_0$ and the hyperfine 
interaction $V_{\text{hyp}}$ that we present in detail below.

As shown in~\cite{ChJ98}, we 
use the
variational principle for the ground state wave function 
in Eq.~(\ref{E:phi_1}) to determine the value of the 
variational parameter $\beta$ by satisfying
\begin{equation} \label{E:variational_principle}
  \frac{\partial\left<\phi_{1S}\left|\left(
    H_0+V_0
  \right)\right|\phi_{1S}\right>}{\partial\beta}=0
,\end{equation}
where $V_0$ is the interacting potential consisting of 
(a) Coulomb
plus HO, and (b) Coulomb plus linear potential
\begin{equation}
  V_0(r)=a+V_{\text{conf}}(r)-\frac{4\kappa}{3r}
.\end{equation}
Here, $V_{\text{conf}}(r)=br^2$ 
[$V_{\text{conf}}(r)=br$] for 
HO [linear]
confining potentials. Once the optimal parameter $\beta$ for
the ground state is determined, we use the same $\beta$ 
for the first and second excited states. This
ensures that all these states are orthogonal.

To distinguish the vector meson from 
the pseudoscalar meson, we include a hyperfine 
interaction~\cite{GoI85,ScI95}
\begin{equation} \label{E:V_hyp_new}
  V_{\text{hyp}}(r)
  =
  \sqrt{\frac{m_qm_{\bar{q}}}{E_qE_{\bar{q}}}}
  \frac{32\pi\kappa\vec{S}_q\cdot\vec{S}_{\bar{q}}}
       {9m_qm_{\bar{q}}}
       \delta^3(\vec{r})
  \sqrt{\frac{m_qm_{\bar{q}}}{E_qE_{\bar{q}}}}
,\end{equation}
where $E_q=\sqrt{m_q^2+\vec{k}^2}$, 
$E_{\bar{q}}=\sqrt{m_{\bar{q}}^2+\vec{k}^2}$
and $\left<\vec{S}_q\cdot\vec{S}_{\bar{q}}\right>=1/4~[-3/4]$
for vector [pseudoscalar] mesons.
Since we are dealing with
light mesons including radially excited states, 
this relativistic correction
is essential. 
If we were to calculate the mass splitting 
${\Delta}M_{nS}$ 
between pseudoscalar and 
vector mesons using the non-relativistic
$V_{\text{hyp}}$ given by
\begin{equation} \label{E:V_hyp_old}
  V_{\text{hyp}}(r)
    =\frac{2\vec{S}_q\cdot\vec{S}_{\bar{q}}}{3m_qm_{\bar{q}}}
       \nabla^2V_{\text{Coul}}
    =\frac{32\pi\kappa\vec{S}_q\cdot\vec{S}_{\bar{q}}}
          {9m_qm_{\bar{q}}}
       \delta^3(\vec{r})          
\end{equation}
then one gets
\begin{equation}
  \Delta M_{1S} : \Delta M_{2S} : \Delta M_{3S} = 8 : 12 : 15
,\end{equation}
where
\begin{equation}
  \Delta M_{nS}=
    \left<\phi_{nS}\left|
      \frac{32\pi\kappa}{9m_qm_{\bar{q}}}\delta^3(\vec{r})
    \right|\phi_{nS}\right>
.\end{equation}
This result clearly contradicts with the experimental values
$\Delta M_{1S} = 0.630~\text{GeV}$, 
$\Delta M_{2S} \approx 0.100~\text{GeV}$ and 
$\Delta M_{3S} \approx 0.100~\text{GeV}$.

To determine the potential parameters, we proceeded as
follows. 
First, we chose the quark mass $m_q=m_{\bar{q}}$
as an input parameter assuming $m_u=m_d$. For both HO and
linear potentials, we tried reasonable CQ
masses in the range
$0.150~\text{GeV}\lesssim m_q \lesssim 0.300~\text{GeV}$.
The potential parameters $a$, $b$ and $\kappa$ were chosen
to provide an optimal fit for 
$\pi$, $\pi(2S)$, $\rho$ and $\rho(2S)$. 
Then, we predicted the 
meson masses for
$\pi(3S)$ 
and $\rho(3S)$. Our results are summarized in 
Table~\ref{T:mass_spectrum}. As one can see, the results
for a wide range of the CQ mass $m_q$ are 
within the experimental limits of the mass spectrum.
Furthermore, the difference between the results of HO and linear 
potential is quite small once the best fit parameters
are chosen.
Note that
the identity of the higher resonances is not 
completely clear yet~\cite{Cas98}.
While there are some experimental evidences for
rho meson with mass 
$1700\pm20~\text{MeV}$,  
there is also a
reported resonance at 
$2149\pm17~\text{MeV}$. 
In fact, $\rho(1700)$ might be a $3S$-hybrid mixture.
Further consideration should be made to see if
our simple model may or may not be suitable to reproduce
these higher resonances. 

We have also examined 
the Gaussian smearing function to weaken the
singularity of the Delta function in the hyperfine 
interaction
\begin{equation}
  \delta^3(\vec{r})\longrightarrow
  \frac{\sigma^3}{\pi^{3/2}}\exp(-\sigma^2r^2) 
.\end{equation}
It turns out, however, that the smearing effect is 
negligible in our model calculations and the
well known
value of $\sigma=1.8$~\cite{GoI85} did not change our
results appreciably.

The radial wave function in Eq.~(\ref{E:phi_1}) 
has been used successfully
to approximate
the ground state wave function in a couple of papers 
as mentioned above. 
In order to make sure if 
the first and second excited states can be well
approximated with Eqs.~(\ref{E:phi_2}) and 
(\ref{E:phi_3}),
we varied the wave function
of the first excited state to the mixed wave function
\begin{equation} \label{E:mixed_2S}
  \widetilde{\left|2S\right>}
    =f_2\left|2S\right>+f_3\left|3S\right>
,\end{equation}
where $\left|2S\right>$ and $\left|3S\right>$
denote the second and third
HO states.
We determined the parameters $f_2$ and $f_3$ using the 
variational principle in 
Eq.~(\ref{E:variational_principle})
as $f_2=0.982$ and $f_3=0.190$, 
showing that the added term is much suppressed.
For the second excited state we used the 
wave function
\begin{equation}
  \widetilde{\left|3S\right>}
    =f_3\left|2S\right>-f_2\left|3S\right>
\end{equation}
which is orthogonal to Eq.~(\ref{E:mixed_2S}).
However,
the calculated mass eigenvalues for these states
deviate less than
$2\%$ from our values determined by using 
Eqs.~(\ref{E:phi_2}) 
and (\ref{E:phi_3}). Therefore, we trust that our
approximation of using Eqs.~(\ref{E:phi_2}) 
and (\ref{E:phi_3}) 
is well justified.

While all the parameter sets summarized in 
Table~\ref{T:mass_spectrum} give a good
agreement with the mass spectrum,
the pion decay constant is rather sensitive 
to the choice of $m_q$.
Following the LC approach of~\cite{Jau90,Jau91} 
for the pion decay constant 
given by
\begin{equation}
  \left<0\left|
    \bar{q}\gamma^{+}\gamma_5q
  \right|\vec{P},00\right>\sqrt{2P^{+}}
  =iP^{+}\sqrt{2}f_\pi
,\end{equation}
one gets
\begin{equation}
  f_\pi=\frac{\sqrt{6}}{(2\pi)^{3/2}}
        \int d^2\vec{k}_{\perp}\int dx
        \sqrt{\frac{\partial k_n}{\partial x}}
        \frac{m_q}{\sqrt{m_q^2+\vec{k}_{\perp}^2}}
        \phi_{1S}(x, \vec{k}_{\perp})
.\end{equation}
As shown in Table~\ref{T:mass_spectrum},
the calculated value of $91.9~\text{MeV}$ 
for 
$m_q=0.190~\text{GeV}$ and the HO potential 
is in good agreement with the 
experimental value of $92.4\pm0.25~\text{MeV}$ 
from~\cite{Cas98}.
We therefore from now on use the 
parametrization where $m_q=0.190~\text{GeV}$ and 
$\beta=0.4957~\text{GeV}$. 
In our work, we don't need 
to introduce an
arbitrary axial-vector coupling constant on the level of 
CQ~\cite{CGN94}.

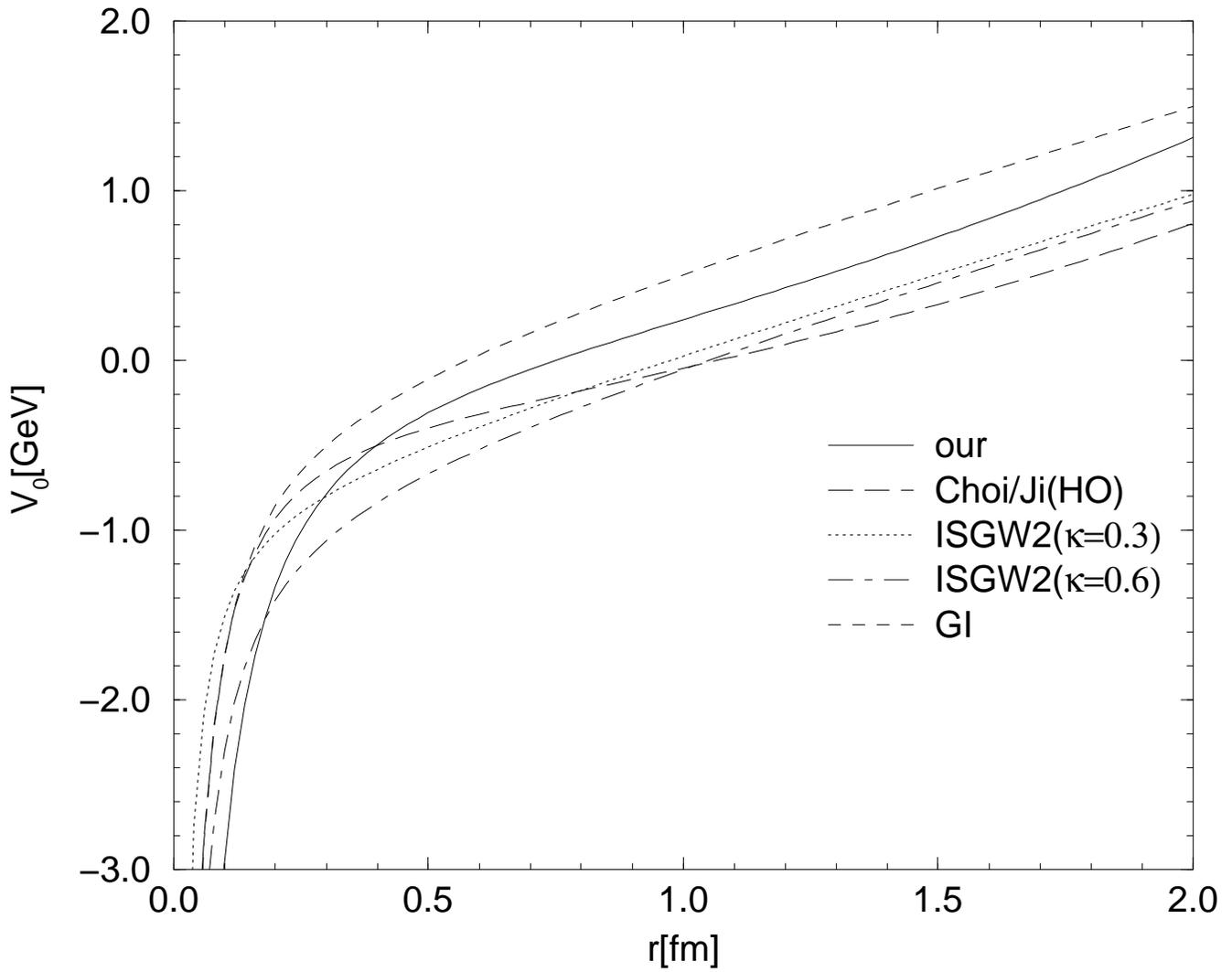
\begin{figure}
  \caption{The meson interaction potential $V_0(r)$. 
           Our parameters are compared with the HO 
           potential of~\protect\cite{ChJ98}, the 
           quasi-relativistic
           potential of ISGW2 ($\kappa=0.3$ and $\kappa=0.6$)
           ~\protect\cite{ScI95} and the relativistic potential of
           GI~\protect\cite{GoI85}.
           }
  \label{F:potential}
\end{figure}

In Fig.~\ref{F:potential}, we compare our 
central potential with those of other 
publications~\cite{GoI85,ScI95,ChJ98}
in the 
interesting range of up to $2~\text{fm}$.
Our potential 
seems quite comparable to the other calculations.

\subsection{CQ Form Factor Parameters}
\label{S:CQ_form_factor_parameters}
While the CQ were often treated as pointlike particles,
there are hints in the  
literature~\cite{CGN94,CGN95b,CGN95c,CGN95d,CGN95a} 
that
the CQ may not be treated as pointlike particles.
Especially, the Gerasimov sum-rule 
calculation~\cite{Ger95}
indicates that the CQ should be
treated as extended particles. Thus, we introduced
Dirac and Pauli form factors on the level of CQ. 
In fact, the Gaussian parameter $\beta=0.4957~\text{GeV}$ 
of our parametrization is
somewhat larger than that of other 
calculations~\cite{ChJ98,Jau90,Jau91,ScI95} 
where $\beta$ is generally
in the range $\beta\approx0.3\ldots 0.4~\text{GeV}$. 
Therefore, it seems important to take into account these
CQ form factors in order to
obtain comparable results with the 
experimental data for the hadronic properties.

We substitute $j_\mu=e_q\gamma_\mu$ at a quark-photon
coupling by the more general form
\begin{equation} \label{J_mu}
  {J_q}_\mu
  =
  F_D^{(q)}(Q^2)e_q\gamma_\mu
  +F_P^{(q)}(Q^2)\kappa_qi\sigma_{\mu\nu}
    \frac{q^\nu}{2m_q}
,\end{equation}
where the Dirac and Pauli form factors on the level of CQ
are normalized as $F_D^{(q)}(0)=F_P^{(q)}(0)=1$. 
Here, $e_q$ and $\kappa_q$ are the CQ charge and 
anomalous magnetic moment,
$Q^2=-q^2$ is the 4-momentum transfer square and
$\sigma_{\mu\nu}=\frac{i}{2}[\gamma_\mu,\gamma_\nu]$.
For the Dirac and Pauli CQ form factors 
$F_D$ and $F_P$, we adopt
simple monopole and dipole forms
\begin{equation} \label{E:CQ_form_factors}
  F_D^{(q)}(Q^2)=\frac{1}{1+<{r_D^{(q)}}^2>Q^2/6}
  \quad\text{and}\quad
  F_P^{(q)}(Q^2)
    =\frac{1}{\left(1+<{r_P^{(q)}}^2>Q^2/12\right)^2}
,\end{equation}
as other authors~\cite{CGN95c} used.
It turns out
to be sufficient for our purpose to use
this simple monopole and dipole
form.

To fix the model parameters 
$<r_D^2>$, $<r_P^2>$,
$\kappa_u$ and 
$\kappa_d$, we fit our model to the available 
pion form factor data and to the experimental values
of the transition magnetic 
moments $\mu_{\rho\pi}$ and $\mu_{\omega\pi}$.
The details of the fitting procedure are presented
in the Appendix.

Our determined values for the parameters $<r_D^2>$,
$<r_P^2>$, $\kappa_u$ and $\kappa_d$ are shown 
in Table~\ref{T:parameters}.
Note that these parameters 
are 
comparable to those determined in~\cite{CGN95c}.
Also, we calculated the ratio 
$(e_u+\kappa_u)/(e_d+\kappa_d)$ which is predicted 
in~\cite{Ger95} as $-1.80\pm0.02$, which is comparable
to our value of $-1.94$.

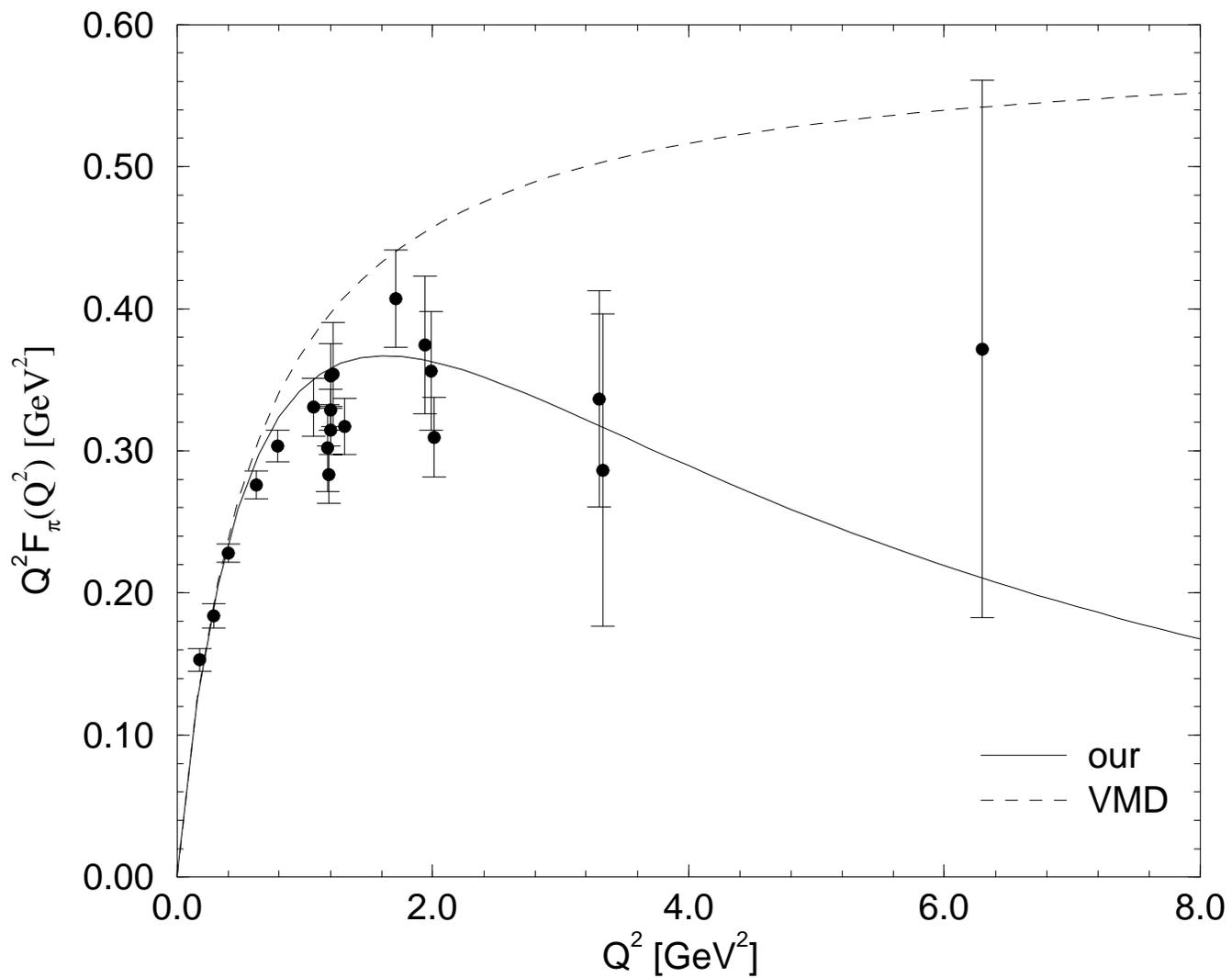
\begin{figure}
  \caption{The $\pi^{+}$ form factor times $Q^2$. We also
           show the predictions from a simple VMD model.
           The experimental
           data is taken from~\protect\cite{BBH78}}
  \label{F:ff_pi}
\end{figure}

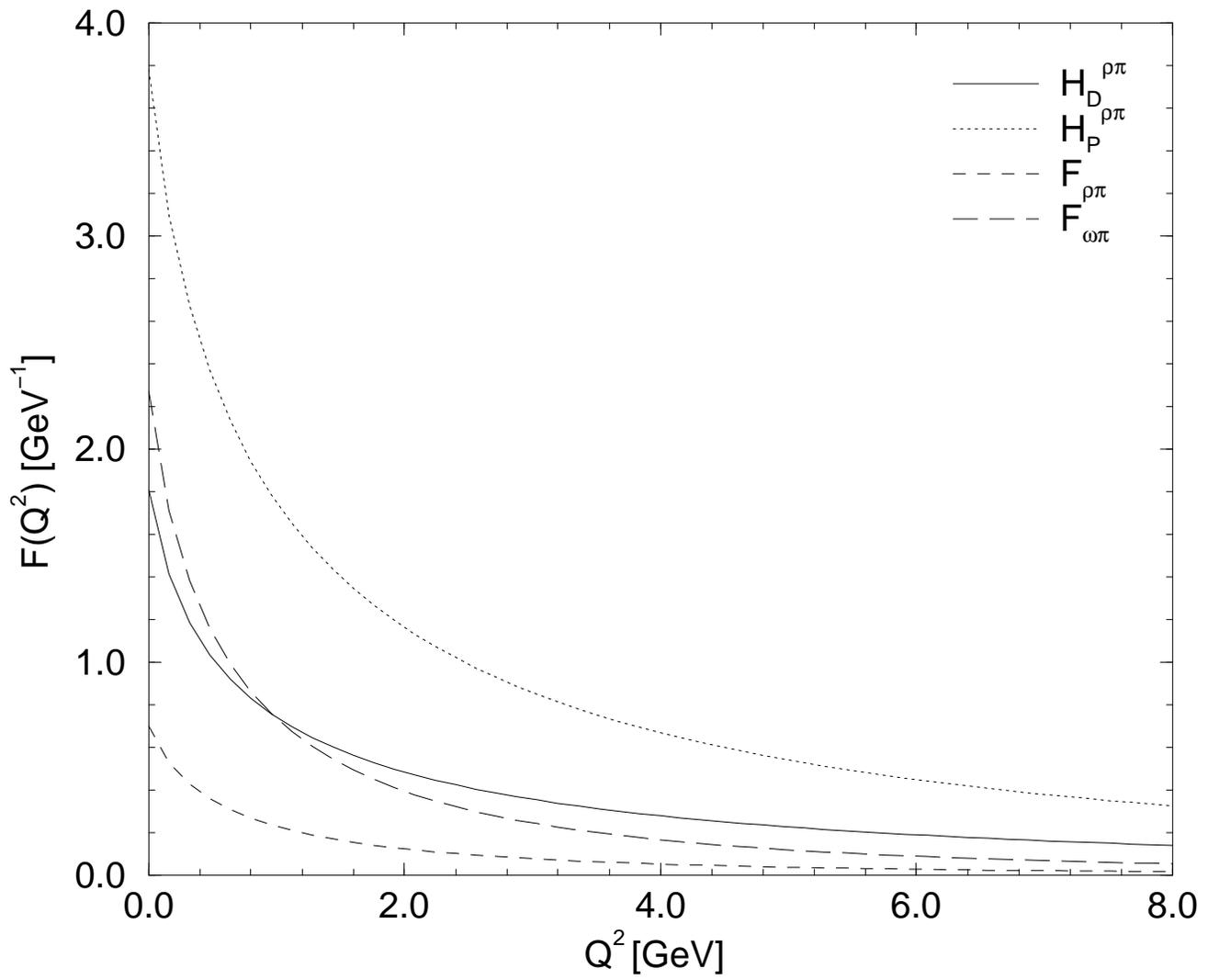
\begin{figure}
  \caption{The radiative
           $\rho^{+}\to\pi^{+}\gamma^{*}$ and
           $\omega\to\pi^{0}\gamma^{*}$ transition
           form factors $F_{\rho\pi}$ and $F_{\omega\pi}$. 
           We also show the body form factors
           $H_D^{\rho\pi}$ and $H_P^{\rho\pi}$}
  \label{F:ff_rho2pi}
\end{figure}

In Fig.~\ref{F:ff_pi}, we show our calculation for 
$F_{\pi}$ and compare it to the experimental data taken
from~\cite{BBH78}.  
We also compare our calculation to a simple 
Vector Meson Dominance (VMD) model where
$F_\pi^{\text{VMD}}(Q^2)=1/(1+Q^2/M_\rho^2)$.
In Fig.~\ref{F:ff_rho2pi}, we show 
the radiative transition
form factors $F_{\rho\pi}$ and $F_{\omega\pi}$
and the body form factors
$H_D^{\rho\pi}$ and $H_P^{\rho\pi}$.

\section{Calculation of Hadronic Properties} 
\label{S:hadronic_properties}
Having fixed all the parameters of our model in 
Sec.~\ref{S:fixing_model_parameters}, 
we now present our predictions of
various hadronic properties of pseudoscalar and vector mesons
including also the radially excited states.

\subsection{$\rho^{0}$ Decay Constant $f_\rho$}
From the definition~\cite{Jau91}
\begin{equation}
  <0|{J_q}_{\mu}|\vec{P},1J_3>\sqrt{2P^{+}}
    =\epsilon_{\mu}(J_3)M_\rho f_{\rho}
,\end{equation}
we calculate the rho decay constant 
$f_\rho$ including both Dirac and Pauli 
form factors on the level of CQ by 
taking ${J_q}_{\mu}$ as defined in Eq.~(\ref{J_mu}).
We obtain for $\rho^0=(u\bar{u}-d\bar{d})/\sqrt{2}$
\begin{equation}
  f_\rho
    =\frac{1}{\sqrt{2}}
     \left(
       [e_u-e_d]F_D^{(q)}(M_\rho^2)I_D
       +[\kappa_u-\kappa_d]F_P^{(q)}(M_\rho^2)I_P
     \right)
,\end{equation}
where
\begin{equation}
  I_D
    =\frac{\sqrt{3}}{(2\pi)^{3/2}}
     \int d^2\vec{k}_{\perp} \int dx
     \frac{1}{x(1-x)}\frac{1}{\sqrt{M_0}}
     \left(m+\frac{2\vec{k}_{\perp}^2}{\lambda}\right)
     \phi_{1S}(x,\vec{k}_{\perp})
\end{equation}
and
\begin{equation}
  I_P
    =\frac{\sqrt{3}}{4(2\pi)^{3/2}m_q}
     \int d^2\vec{k}_{\perp} \int dx
     \frac{\sqrt{M_0}}{x(1-x)}
     \left(\frac{4\vec{k}_{\perp}^2}{\lambda}-M_0\right)
     \phi_{1S}(x,\vec{k}_{\perp})  
.\end{equation}
Note here that the momentum transfer square
equals the mass 
of the rho meson
$Q^2=M_\rho^2=(0.770~\text{GeV})^2$. While $I_D$ has been 
already derived in~\cite{Jau91}, $I_P$ is our new body
form
factor related to the Pauli form factor in the electromagnetic
current operator ${J_q}_{\mu}$. Using our parametrization 
(Table~\ref{T:parameters}) we get the value 
$f_\rho=153~\text{MeV}$ which is in a good agreement
with the
experimental data $f_\rho=152.8\pm3.6~\text{MeV}$ 
obtained from the width 
$\Gamma(\rho\to e^{+}e^{-})$~\cite{Cas98}.

\subsection{$\rho^+$ Form Factors}
Our calculation for the $\rho^+$ form factors follows the one 
presented in~\cite{ChJ97}. However,
we again treat the CQ as extended objects characterized 
by Dirac and Pauli form factors in contrast to a treatment as
point like ones.

In the standard LC frame the charge, magnetic and 
quadrupole form factors of a meson
can be obtained from the plus component of three helicity
matrix elements~\cite{BrH92}
\begin{eqnarray}
  F_C&=&
    \frac{1}{(2\alpha+1)}
    \left[
      \frac{16}{3}\alpha\frac{F_{+0}^{+}}{\sqrt{2\alpha}}
      -\frac{1}{3}(2\alpha-3)F_{00}^{+}
      +\frac{2}{3}(2\alpha-1)F_{+-}^{+}
    \right] \\
  F_M&=&
    \frac{2}{(2\alpha+1)}
    \left[
      (2\alpha-1)\frac{F_{+0}^{+}}{\sqrt{2\alpha}}
      +F_{00}^{+}-F_{+-}^{+}
    \right] \\
  F_Q&=&
    \frac{1}{(2\alpha+1)}
    \left[
      2\frac{F_{+0}^{+}}{\sqrt{2\alpha}}
      -F_{00}^{+}
      -\frac{\alpha+1}{\alpha}F_{+-}^{+}
    \right]
\end{eqnarray}
where 
$F_{\lambda'\lambda}^\mu=
   \left<P',\lambda'\left|
     \left(J_u^\mu-J_d^\mu\right)
   \right|P,\lambda\right>$, 
$\alpha=\frac{Q^2}{4M_\rho^2}$ is a kinematic factor and 
$J_q^\mu$ is defined in Eq.~(\ref{J_mu}).
This representation is not unique; there are different 
prescriptions in the 
literature, as discussed 
in~\cite{CGN95a}. This ambiguity is reflected in the 
fact that the angular condition
\begin{equation} \label{E:angular_condition}
  \Delta(Q^2)
    =(1+2\alpha)F_{++}^{+}+F_{+-}^{+}
     -\sqrt{8\alpha}F_{+0}^{+}-F_{00}^{+}
    =0
\end{equation}
is in general violated unless the exact Poincar\'{e} covariant 
current operator beyond one-body sector is used.

At zero momentum transfer, these form factors are proportional 
to the meson charge $e$, 
magnetic moment $\mu_1$ and quadrupole moment $Q_1$:
\begin{equation}
  F_C(0)=1, \quad\quad
  eF_M(0)=2M_\rho \mu_1 \quad\quad \text{and} \quad\quad
  eF_Q(0)=M_\rho^2 Q_1
.\end{equation}

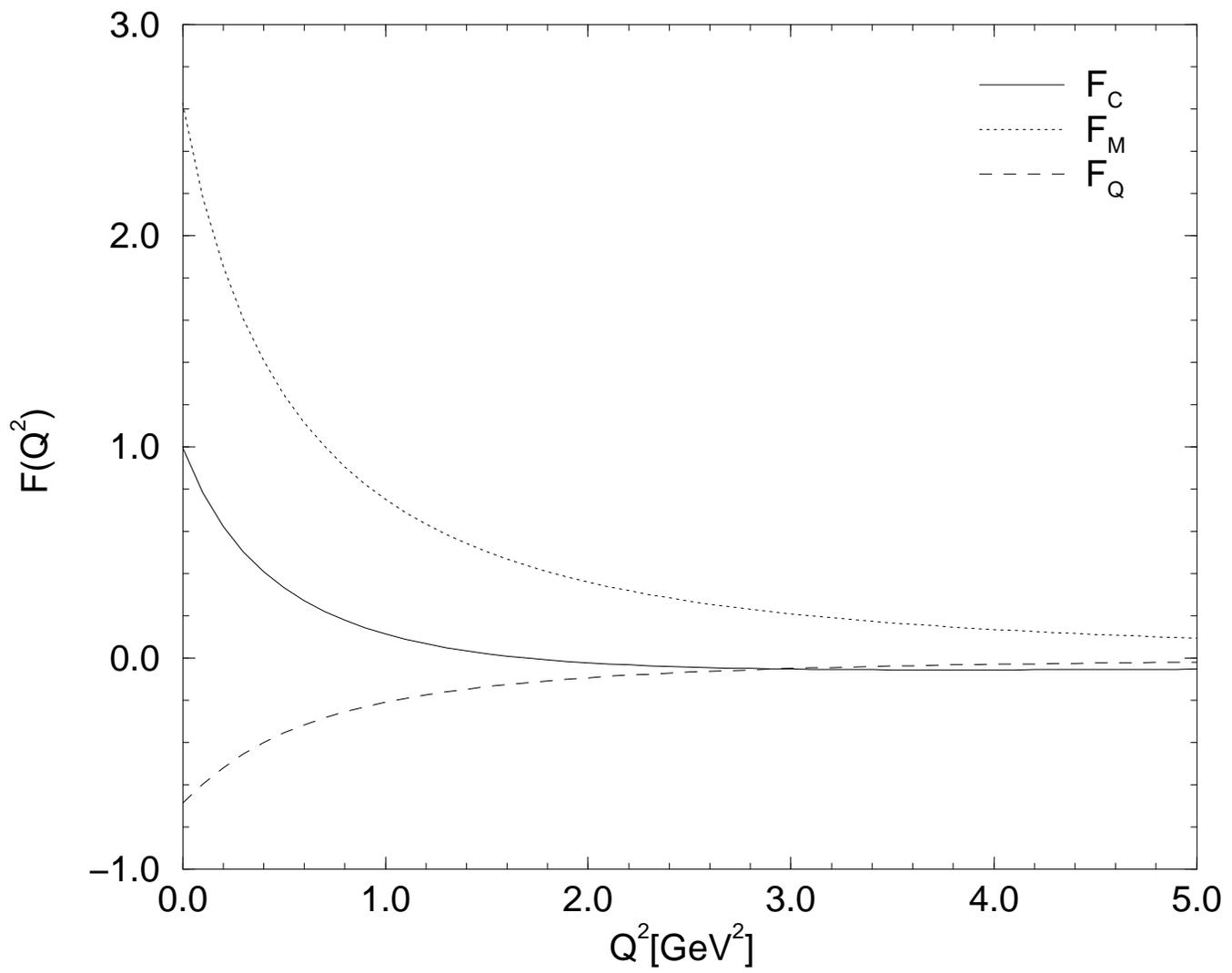
\begin{figure}
  \caption{The rho form factors including both Dirac and Pauli
           form factors on the level of CQ.}
  \label{F:ff_rho}
\end{figure}

In the LC quark model, the matrix element 
$\left<P',\lambda'\left|J_q^\mu\right|P,\lambda\right>$ 
can be calculated by the 
convolution of initial and final LC wave function of
a meson
\begin{eqnarray}
  \left<P',\lambda'\left|J_q^\mu\right|P,\lambda\right>
  =&&
  \sum_{\lambda_q,\lambda_{q'},\lambda_{\bar{q}}}
  \int d^2\vec{k}_{\perp} \int dx
  \sqrt{\frac{\partial k_n}{\partial x}}
  \sqrt{\frac{\partial k'_n}{\partial x}}
  \phi_{1S}(x, \vec{k}_{\perp})
  \phi^*_{1S}(x, \vec{k}_{\perp}') \nonumber \\
    &&    
    \times
    \left[
      \frac{\bar{u}(k_{q'},\lambda_{q'})}{\sqrt{{k_{q'}^+}}}
      \frac{J_q^\mu}{2}
      \frac{u(k_q,\lambda_q)}{\sqrt{{k_q}^{+}}}
    \right]
    {\mathcal R}'^*
      (\lambda_{q'},\lambda_{\bar{q}};1,\lambda')
    {\mathcal R}(\lambda_q,\lambda_{\bar{q}};1,\lambda)
.\end{eqnarray}
After a straightforward calculation,
choosing the $+$~component of the current, 
we get 
\begin{equation}
  F^+_{\lambda'\lambda}(Q^2)
  =
  (e_u-e_d)F_D^{(q)}(Q^2)I_D^{\lambda'\lambda}(Q^2)
  +(\kappa_u-\kappa_d)F_P^{(q)}(Q^2)I_P^{\lambda'\lambda}(Q^2) 
,\end{equation}
where the body form
factors related to the Dirac part are given by
\begin{eqnarray}
  I_D^{00}(Q^2)=&&
   \int d^2\vec{k}_{\perp} \int dx
   \sqrt{\frac{\partial k_n}{\partial x}}
   \sqrt{\frac{\partial k_n'}{\partial x}}
   \phi_{1S}(x,\vec{k}_{\perp})
   \phi^*_{1S}(x,\vec{k}_{\perp}') 
   \frac{1}{x(1-x)\lambda\lambda'} \nonumber \\
   &&
   \times
   \left[
     (1-2x)^2\vec{k}_{\perp}\vec{k}_{\perp}'
     +\left(2x[1-x]M_0'+m\right)\left(2x[1-x]M_0+m\right)
   \right]
,\end{eqnarray} 
\begin{eqnarray}
  I_D^{+0}(Q^2)=&&
   \int d^2\vec{k}_{\perp} \int dx
   \sqrt{\frac{\partial k_n}{\partial x}}
   \sqrt{\frac{\partial k_n'}{\partial x}}
   \phi_{1S}(x,\vec{k}_{\perp})
   \phi^*_{1S}(x,\vec{k}_{\perp}')
   \frac{1-2x}{\sqrt{2}M_0'x\lambda\lambda'} \nonumber \\
   &&
   \times
   \left[
     2xk_xM_0'(M_0'-M_0)
     +Q\left(2k_y^2-2x(1-x)M_0'M_0-M_0'm\right)
   \right]
,\end{eqnarray} 
\begin{eqnarray}
  I_D^{+-}(Q^2)=&&
   \int d^2\vec{k}_{\perp} \int dx
   \sqrt{\frac{\partial k_n}{\partial x}}
   \sqrt{\frac{\partial k_n'}{\partial x}}
   \phi_{1S}(x,\vec{k}_{\perp})
   \phi^*_{1S}(x,\vec{k}_{\perp}')
   \frac{1}{M_0M_0'x(1-x)\lambda\lambda'} \nonumber \\ 
   &&
   \times
   \left[
     \left(mM_0+mM_0'+2m^2+2x(1-x)M_0M_0'\right)k_xk_x'
     +\left(2m^2-2x[1-x]M_0M_0'\right)k_y^2 
                     \right. \nonumber \\
                     &&   \left.
     -2{k_x}^2{k_x'}^2
     -m\left(\lambda'k_x^2+\lambda{k_x'}^2\right)
     +2k_y^4
   \right]
\end{eqnarray} 
and
\begin{eqnarray}
  I_D^{++}(Q^2)=&&
   \int d^2\vec{k}_{\perp} \int dx
   \sqrt{\frac{\partial k_n}{\partial x}}
   \sqrt{\frac{\partial k_n'}{\partial x}}
   \phi_{1S}(x,\vec{k}_{\perp})
   \phi^*_{1S}(x,\vec{k}_{\perp}')
   \frac{1}{M_0M_0'x(1-x)\lambda\lambda'} \nonumber \\
   &&
   \times
   \left[
     \left({k'}_{\perp}^2+m\lambda'\right)
         \left(k_{\perp}^2+m\lambda\right)
     +\left(\vec{k}_{\perp}'\vec{k}_{\perp}\right)^2
     -(1-x)^2Q^2k_y^2   
                 \right. \nonumber \\
                 &&    \left.
     +\left(2m^2+mM_0+mM_0'+[2x^2-2x+1]M_0M_0'\right)
       \vec{k}_{\perp}'\vec{k}_{\perp}  
   \right]
.\end{eqnarray} 
Here, $\lambda'=2m+M_0'$.
These formulas have already been used by the authors 
of~\cite{CGN95a} where they calculated the $\rho^+$ form factors
including only the Dirac part of the CQ form factors by
assuming that the anomalous magnetic moments of the CQ
are negligible. We, however, present a complete calculation 
including also the Pauli form factor part. For the new 
body form factors related to the Pauli part, we obtain
\begin{eqnarray}
  I_P^{00}(Q^2)=&&
   \int d^2\vec{k}_{\perp} \int dx
   \sqrt{\frac{\partial k_n}{\partial x}}
   \sqrt{\frac{\partial k_n'}{\partial x}}
   \phi_{1S}(x,\vec{k}_{\perp})
   \phi^*_{1S}(x,\vec{k}_{\perp}')
   \frac{Q(1-2x)}{2x(1-x)m\lambda\lambda'} \nonumber \\
   &&
   \times
   \left[
     \frac{m\lambda'+2{k_{\perp}'}^2}{M_0'}k_x
     -\frac{m\lambda+2k_{\perp}^2}{M_0}k_x'
   \right]
,\end{eqnarray} 
\begin{eqnarray}
  I_P^{+0}(Q^2)=&&
   \int d^2\vec{k}_{\perp} \int dx
   \sqrt{\frac{\partial k_n}{\partial x}}
   \sqrt{\frac{\partial k_n'}{\partial x}}
   \phi_{1S}(x,\vec{k}_{\perp})
   \phi^*_{1S}(x,\vec{k}_{\perp}')
   \frac{Q}{2\sqrt{2}x(1-x)m\lambda\lambda'} \nonumber \\
   &&
   \times
   \left[
     \frac{1-2x}{M_0'}
     \left(
       [m+xM_0']\vec{k'}_{\perp}\vec{k}_{\perp}
       -[m+(1-x)M_0'][k_x'k_x-k_y^2]
     \right)
             \right. \nonumber \\
             && \left.
     -\frac{1}{M_0'M_0}
     \left(
       [\lambda m+2k_{\perp}^2][{k_x'}^2-k_y^2]
       +[\lambda'm+{k_{\perp}'}^2]
               [\lambda m+2k_{\perp}^2]
     \right)
   \right]
,\end{eqnarray} 
\begin{eqnarray}
  I_P^{+-}(Q^2)=&&
   \int d^2\vec{k}_{\perp} \int dx
   \sqrt{\frac{\partial k_n}{\partial x}}
   \sqrt{\frac{\partial k_n'}{\partial x}}
   \phi_{1S}(x,\vec{k}_{\perp})
   \phi^*_{1S}(x,\vec{k}_{\perp}')
   \frac{Q}{2x(1-x)m\lambda\lambda'M_0M_0'} \nonumber \\
   &&
   \times
   \left[
     \left(m+[1-x]M_0'\right)
       \left(k_x'\left[k_x^2-k_y^2\right]-2k_xk_y^2\right)
     +\left({k_{\perp}'}^2+\lambda'm\right)(m+xM_0)k_x
             \right. \nonumber \\ 
             && \left.
     -(m+[1-x]M_0)
       \left(k_x\left[{k_x'}^2-k_y^2\right]-2k_x'k_y^2\right)
     -\left(k_{\perp}^2+\lambda m\right)
       \left(m+xM_0'\right)k_x'
   \right]
\end{eqnarray} 
and
\begin{eqnarray}
  I_P^{++}(Q^2)=&&
   \int d^2\vec{k}_{\perp} \int dx
   \sqrt{\frac{\partial k_n}{\partial x}}
   \sqrt{\frac{\partial k_n'}{\partial x}}
   \phi_{1S}(x,\vec{k}_{\perp})
   \phi^*_{1S}(x,\vec{k}_{\perp}')
   \frac{Q}{2x(1-x)m\lambda\lambda'M_0M_0'} \nonumber \\
   &&
   \times
   \left[
     -(m+(1-x)M_0')\left(k_{\perp}^2+\lambda m\right)k_x'
     -\left(m+xM_0\right)
      \left(k_x\left[{k_x'}^2-k_y^2\right)+2k_x'k_y^2\right)
                 \right. \nonumber \\
                 && \left.
     +(m+(1-x)M_0)\left({k_{\perp}'}^2+\lambda'm\right)k_x
     +(m+xM_0')
       \left(k_x'\left[k_x^2-k_y^2\right]+2k_xk_y^2\right)
   \right]
.\end{eqnarray} 
We show the result for the $\rho^+$
form factors using our parametrization
(Table~\ref{T:parameters}) in Fig.~\ref{F:ff_rho}.  
In Table~\ref{T:observables}, we list our calculations
for $\mu_1$ and $Q_1$. Note that our result is quite 
comparable to other 
calculations~\cite{BaE85,CGN95a,ChJ97}. 

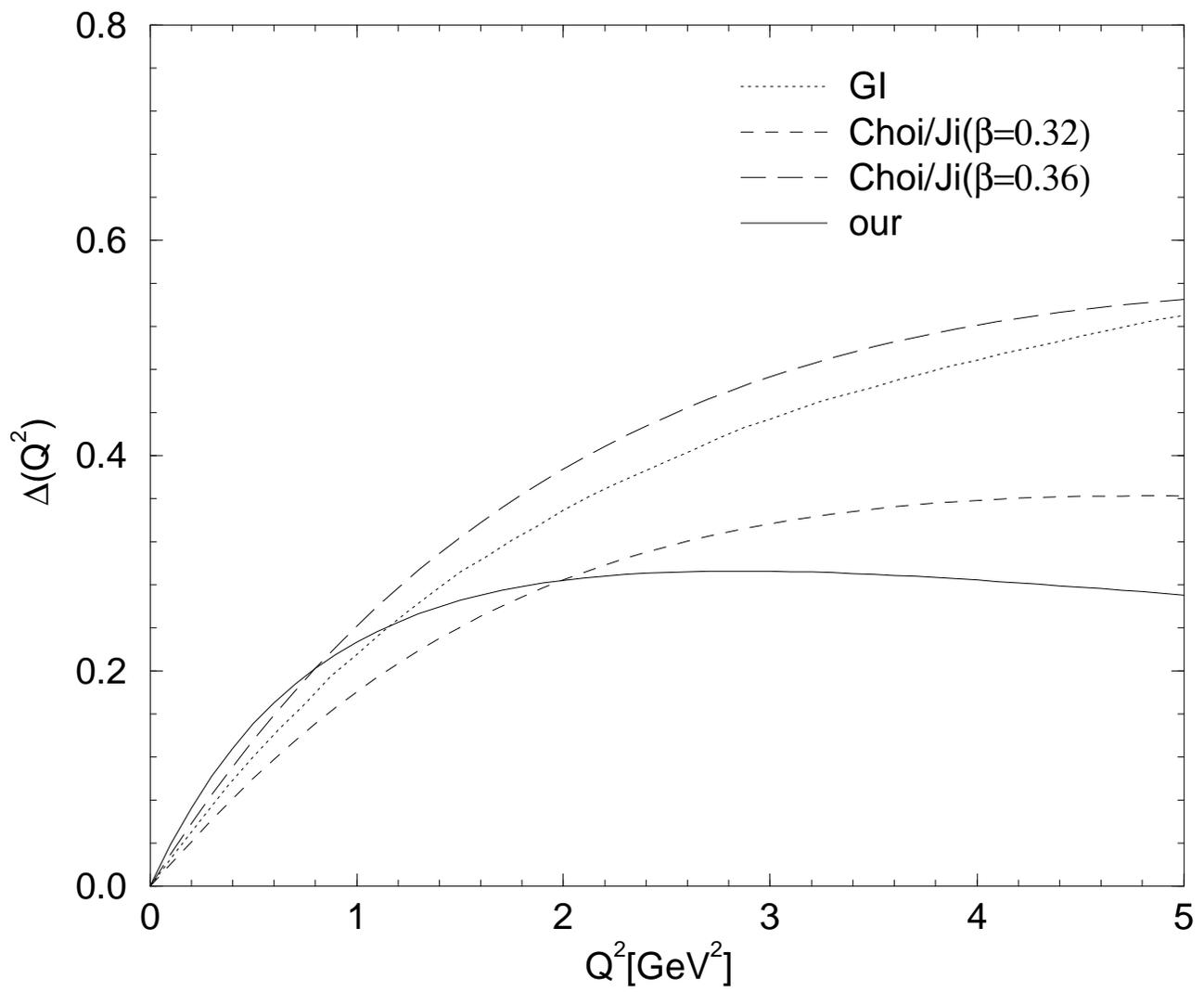
\begin{figure}
  \caption{The angular condition tested by $\Delta(Q^2)$. 
           Our result is compared to other calculations
           ~\protect\cite{ChJ97,CGN95a}.
           A feature of our model is the fact
           that the violation of the 
           angular condition is smaller than in
           other models.}
  \label{F:angular_condition}
\end{figure}

Using Eq.~(\ref{E:angular_condition}), 
we have also calculated the 
violation of the angular condition and compared it
in Fig.~\ref{F:angular_condition} with several
other calculations~\cite{CGN95a,ChJ97}.
The violation of the angular
condition seems to be   
suppressed by taking into account the structure of CQ,
compared with the other calculations.

\subsection{$\pi^{0}\to \gamma^{*}\gamma$ 
            Transition Form Factor} 
The $\pi^{0}\to \gamma^{*}\gamma$ form factor 
$F_{\pi\gamma}$ has been
calculated quite successfully in several models where the
CQ has been treated as a pointlike 
particle~\cite{Jau91,ChJ98}. 
We show the generalized formulation 
including both Dirac and Pauli form factor of the CQ.

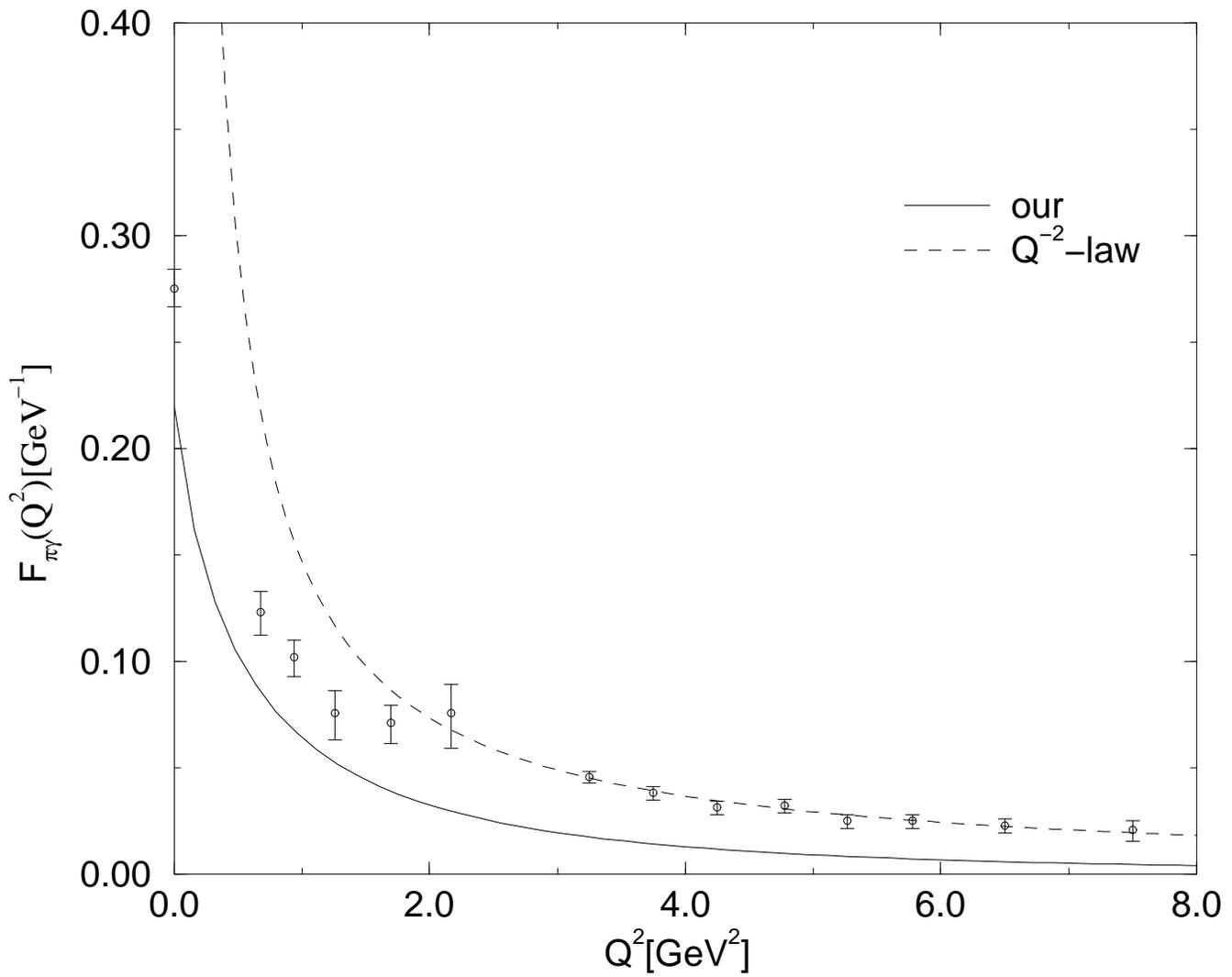
\begin{figure}
  \caption{The $\pi^{0}\to\gamma^{*}\gamma$ form factor
           $F_{\pi\gamma}$. Our 
           result agrees with the experimental data 
           ~\protect\cite{Cel91,Cel95} only for low $Q^2$.
           Using a simple power law Ansatz as suggested
           in~\protect\cite{BJP98} we get agreement for 
           $Q^2>2~\text{GeV}^2$.}
  \label{F:pi_gamma_gamma}
\end{figure}

The $\pi^{0}\to \gamma^{*}\gamma$ transition form factor in 
leading order is defined as
\begin{equation}
  {\Gamma_q}_\mu=
    iG_q(Q^2)
    \epsilon_{\mu\nu\rho\sigma}
    P^\nu\epsilon^\rho {q^{*}}^\sigma
,\end{equation} 
where $P$ is the momentum of the incident pion and $q^{*}$ 
is the momentum of the virtual photon. 
Taking $P^+=1$  
the vertex factor is  
given by 
\begin{eqnarray}
  \Gamma_q^{+}=&&
  \frac{\sqrt{3}}{4(2\pi)^{3/2}}
  \sum_{\lambda, \lambda', \bar{\lambda}}
  \int d^2\vec{k}_{\perp}
  \int dx_1dx_2\delta(1-x_1-x_2)
  \sqrt{\frac{\partial k_n}{\partial x}}
  \phi_{1S}(x,\vec{k}_{\perp})
                   \nonumber \\
                   &&
  \times
  \left[
    {\mathcal R}(\lambda,\bar{\lambda};0,0)
    \frac{\bar{v}_{\bar{\lambda}}(x_2,-\vec{k}_{\perp})}
         {\sqrt{x_2}}
    \epsilon_\mu J^\mu_q
    \frac{u_{\lambda'}(x_1,\vec{k}_{\perp}
                           +\vec{q}_{\perp})}
         {\sqrt{x_1}}
    \frac{\bar{u}_{\lambda'}(x_1,\vec{k}_{\perp}
                           +\vec{q}_{\perp})}
         {\sqrt{x_1}}
    J^{+}_q 
    \frac{u_{\lambda}(x_1,\vec{k}_{\perp})}
         {\sqrt{x_1}}
                  \right. \nonumber \\
                  && \left.
    \frac{1}{q_{\perp}^2
             -\frac{(\vec{k}_{\perp}
                    +\vec{q}_{\perp})^2+m^2}
                   {x_1}
             -\frac{{k_{\perp}^2+m^2}}
                   {x_2}}
    +(x_1\leftrightarrow x_2)
  \right]
.\end{eqnarray}
Note that we have to plug in ${J_q}_\mu$ 
as defined in (\ref{J_mu}) at the
vertices of the real and the virtual photon.
Here, of course,
the momentum transfer square for the real photon 
should be zero.
After a straightforward calculation, we get
\begin{equation} \label{E:F_pi_g_g}
  F_{\pi\gamma}(Q^2)=G_u(Q^2)-G_d(Q^2)
,\end{equation}
where
\begin{eqnarray}
  G_q(Q^2)=&& \label{E:G_q}
  \frac{\sqrt{3}}{\sqrt{2}Q(2\pi)^{3/2}}
  \int d^2\vec{k}_{\perp} \int dx
  \frac{\sqrt{M_0}}{{M_0'}^2x^2(1-x)}
  \phi_{1S}(x,\vec{k}_{\perp})
              \nonumber \\
              &&
  \times
  \left[
    e_qF_D(Q^2)\frac{2}{M_0}
    \left(
      mQe_q-\frac{\kappa_q}{2m}
              \left[
                \left(k_x+2(1-x)Q\right)xM_0^2+(1-x)k_xQ^2
              \right]  
    \right)
           \right. \nonumber \\
               && \left.
    +\kappa_q F_P(Q^2)\frac{1}{m(1-x)M_0}
    \left(
      e_qQ\left[xk_{\perp}^2+m^2\right]
      +e_q[1-x]Q\left[k_x^2-k_y^2\right]
              \right.\right. \nonumber \\
              && \left.\left.
      +e_q[1-x]Q^2k_xe_q
      -\kappa_q\frac{Q}{2}
           \left[k_{\perp}^2+m^2-(1-x)^2Q^2\right]
    \right)
  \right]
.\end{eqnarray}
Our result is shown in Fig.~\ref{F:pi_gamma_gamma}. 
As Jaus~\cite{Jau91} pointed out, the hadron structure 
of the neutral pseudoscalar meson ($\pi^0$)
may not be well enough approximated by the one-loop
calculation.
Gluon-exchange effects may introduce additional structure 
that might lead to a mechanism analogous to the
flavor mixing of isoscalar states.
Especially, at $Q^2=0$, from the one-loop formula
given by Eqs.~(\ref{E:F_pi_g_g}) and (\ref{E:G_q})
and the definition of $\Gamma_{\pi\gamma}$
given by
\begin{equation} \label{E:G_pi_g_g}
  \Gamma_{\pi\gamma}=
  \frac{\pi}{4}\alpha^2
  {F_{\pi\gamma}(0)}^2
  M_{\pi}^3
,\end{equation}
we obtaion 
$\Gamma_{\pi\gamma}=4.97~\text{eV}$.
However, the agreement with the
experimental data
$\Gamma_{\pi\gamma}=7.8\pm0.5~\text{eV}$ 
can be obtained by taking into account the PCAC and
the anomaly of the axial-vector
current~\cite{BeJ69,Adl69},
which predicts
\begin{equation}
  F_{\pi\gamma}(0)=\frac{1}{4\pi^2f_\pi}
.\end{equation}
Using this and Eq.~(\ref{E:G_pi_g_g}),
we obtain
$\Gamma_{\pi\gamma}=7.82~\text{eV}$.

On the other hand, the high-momentum transfer region 
($Q^2\gtrsim~\text{few GeV}^2$) is dominated by the
one-loop off-shell quark contribution as evidenced
from the $Q^2$-behaviour 
$\left(\sim\frac{1}{Q^2}\right)$.
Recently~\cite{BJP98}, it has been shown that for high 
momentum transfer region
the transition form factor can be 
obtained by the renormalization scale and scheme
independent PCQD calculation as
\begin{equation}
  F_{\pi\gamma}(Q^2)=
    \frac{2f_\pi}{Q^2}
    \left(
      1-\frac{5}{3}\frac{\alpha_V(e^{-3/2}Q)}{\pi}
    \right)
,\end{equation}
where $\alpha_V(e^{-3/2}Q)/\pi\approx0.12$. Using this 
relation, one can get a good agreement with the 
experimental data for $Q^2\gtrsim2~\text{GeV}^2$,
as shown in Fig.~\ref{F:pi_gamma_gamma}.

\subsection{Predictions for Radially Excited States} 
We also calculate several hadronic properties
of radially excited states
using the same parameters shown in Table~\ref{T:parameters}.
For these calculations, we use the radial wave functions
$\phi_{2S}$ and $\phi_{3S}$ from 
Eqs.~(\ref{E:phi_2}) and (\ref{E:phi_3}), respectively.

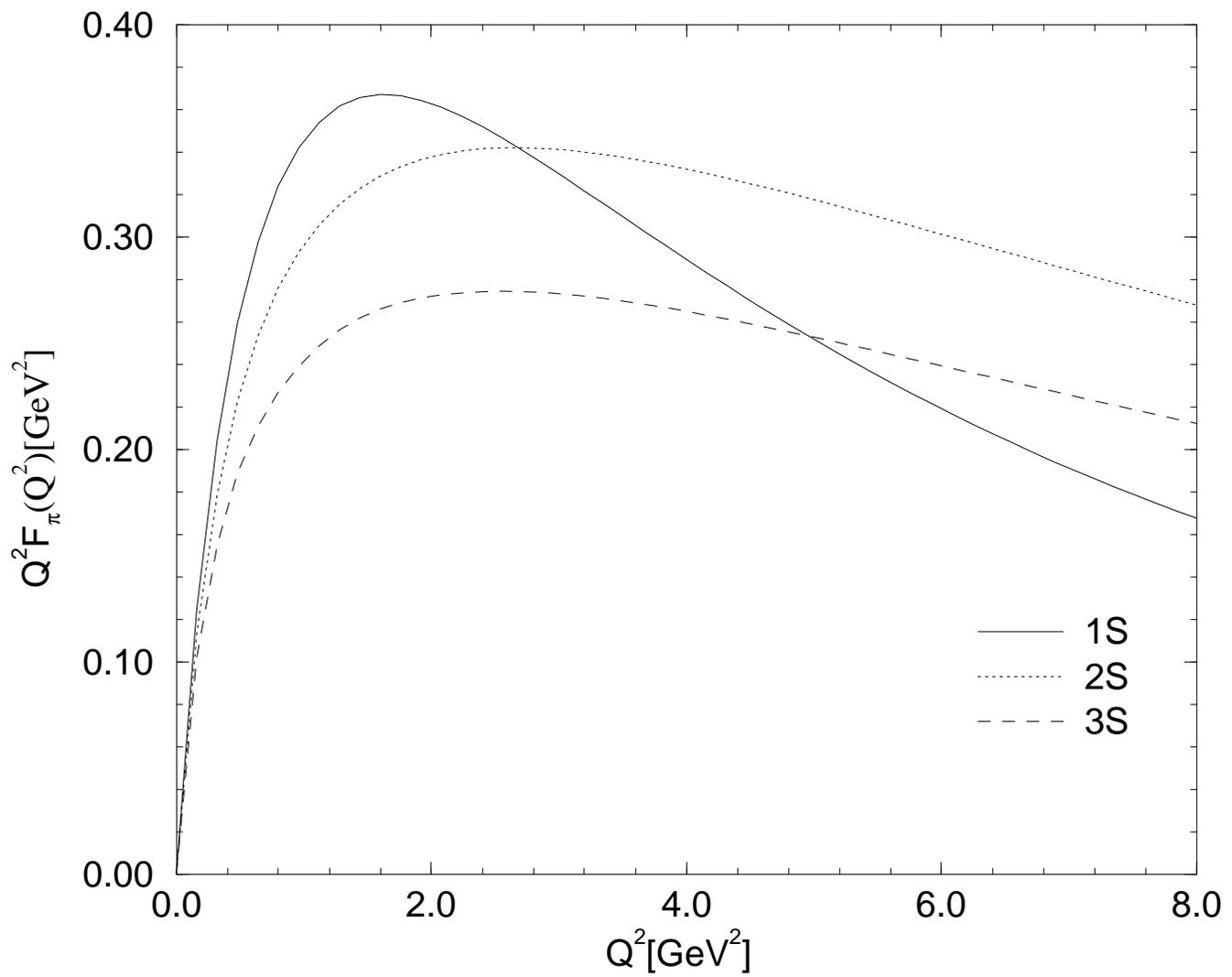
\begin{figure}
  \caption{The $\pi^{+}$ form factor times $Q^2$ for the 
           ground state ($1S$) and the first two radially
           excited states
           ($2S$ and $3S$).}
  \label{F:ff_pi_excited}
\end{figure}

In Table~\ref{T:predictions}, we summarized our results
including the
transition magnetic moments
$\mu_{{\rho'}^+\pi^+}$ and $\mu_{\omega'\pi^0}$, which are
the transitions from
the first exited state to the ground state. 
While the pion and rho decay constants are
proportional to the wave functions at the
origin for the ground state, 
the nodal structures of $2S$ and $3S$ radially
excited states make intuitive predictions on the
ordering of decay constant magnitudes untenable.
We also show our 
result for the
pion form factors for the first and second excited 
states in Fig.~\ref{F:ff_pi_excited}.

\section{Summary and Discussion} 
\label{S:summary}
In this paper, we used a simple relativistic
CQ model to calculate various properties of
radially excited $u$ and
$d$ quark sector mesons. We find that 
a QCD motivated Hamiltonian with a relativistically
corrected hyperfine interaction 
yield the mass spectrum comparable with data.
We also treated the CQ 
not as pointlike objects but as extended ones,
thereby introducing both Dirac and Pauli electromagnetic form
factors for the CQ, and were able to obtain a reasonable
agreement with the data for hadronic properties
like
the pion decay constant $f_\pi$,
the pion form factor $F_\pi$
and the transition magnetic moments $\mu_{\rho\pi}$
and $\mu_{\omega\pi}$.
Furthermore, 
we used this model to calculate 
the rho decay constant $f_\rho$,
the rho form factors
and the $\pi^0\to\gamma^*\gamma$ transition 
form factor $F_{\pi\gamma}$, 
presenting for the first time 
generalized formulas including
both Dirac and Pauli form factors on the level of the CQ. 
We find that the angular condition
$\Delta(Q^2)$ is better satisfied when the CQ form factors
are included.
However, our model 
is intrinsically limited to small momentum 
transfer range.
As an application of our model, we also calculated several
hadronic properties for radially excited states.
Further experimental data on radially excited states 
will give
more stringent test of our model.

\section*{Acknowledgement}
We would like to thank Ho-Meoyng Choi for several
useful discussions.
This work was supported by the Department of Energy under
Grant No.~DE-FG02-96ER40947. 
D.~A. wants to thank the German-American Fulbright Program
and the Friedrich-Ebert-Stiftung for financial support.
The National Energy Research Scientific Computing
Center and the North Carolina Supercomputing Center 
are also acknowledged for the grant of Cray time.

\appendix

\section*{Details of the Determination of the CQ parameters}
The calculation of $F_\pi$ has been
considered in a number of 
references~\cite{CCP88,Dzi88,CGN94,CGN95c}. For 
$\pi^{+}=u\bar{d}$ one gets
\begin{eqnarray}\label{F_pi}
  F_{\pi}(Q^2) 
  &=&
  \left(
    e_uF_D^{(u)}(Q^2)+e_{\bar{d}}F_D^{(\bar{d})}(Q^2)
  \right)H_D^{\pi}(Q^2)
  +\left(\kappa_uF_P^{(u)}(Q^2)
         +\kappa_{\bar{d}}F_P^{(\bar{d})}(Q^2)
   \right)H_P^{\pi}(Q^2) \nonumber \\
  &=&
  F_D^{(q)}(Q^2)H_D^{\pi}
  +\left(\kappa_u-\kappa_d\right)
   F_P^{(q)}(Q^2)H_P^{\pi}(Q^2) 
,\end{eqnarray}
assuming the same internal structure for the $u$
and $d$ CQ, $F_D^{(u)}=F_D^{(d)}\equiv F_D^{(q)}$
and $F_P^{(u)}=F_P^{(d)}\equiv F_P^{(q)}$.
The body form factors $H_D^{\pi}$ and 
$H_P^{\pi}$ 
are given by~\cite{CGN95c}
\begin{equation}
  H_D^{\pi}(Q^2)
  =\int d^2\vec{k}_{\perp} \int dx
   \sqrt{\frac{\partial k_n}{\partial x}}
   \sqrt{\frac{\partial k_n'}{\partial x}}
   \phi_{1S}(x,\vec{k}_{\perp})
   \phi^*_{1S}(x,\vec{k}_{\perp}')
   \frac{m^2+\vec{k}_{\perp}\vec{k}_{\perp}'}
        {x(1-x)M_0M_0'}
\end{equation}
and
\begin{equation}
  H_P^{\pi}(Q^2)
  =\int d^2\vec{k}_{\perp} \int dx
   \sqrt{\frac{\partial k_n}{\partial x}}
   \sqrt{\frac{\partial k_n'}{\partial x}}
   \phi_{1S}(x,\vec{k}_{\perp})
   \phi^*_{1S}(x,\vec{k}_{\perp}')
   \frac{-Q^2}{2xM_0M_0'}
.\end{equation}
$M_0'$ is given by 
${M_0'}^2=\frac{\vec{k'}_{\perp}^2+m_q^2}{x(1-x)}$
and $\vec{k}_{\perp}'$ by 
$\vec{k}_{\perp}'=\vec{k}_{\perp}+(1-x)\vec{q}_{\perp}$.

Following a similar spin and flavour algebra for 
$\rho^+=u\bar{d}$, 
$\omega=1/\sqrt{2}(u\bar{u}+d\bar{d})$ and
$\pi^0=1/\sqrt{2}(u\bar{u}-d\bar{d})$,
one gets for the radiative 
$\rho^+\to\pi^+\gamma^*$ and 
$\omega\to\pi^0\gamma^*$ transition form factors
\begin{eqnarray}\label{F_rho_pi}
  F_{\rho\pi}(Q^2)
  &=&
  \left(
    e_uF_D^{(u)}(Q^2)-e_{\bar{d}}F_D^{(\bar{d})}(Q^2)
  \right)H_D^{\rho\pi}(Q^2)
  +\left(
     \kappa_uF_P^{(u)}(Q^2)
             -\kappa_{\bar{d}}F_P^{(\bar{d})}(Q^2)
   \right)H_P^{\rho\pi}(Q^2) \nonumber \\
  &=&
  \frac{1}{3}F_D^{(q)}H_D^{\rho\pi}(Q^2)
  +\left(\kappa_u+\kappa_d\right)
   F_P^{(q)}(Q^2)H_P^{\rho\pi}(Q^2) 
\end{eqnarray}
and
\begin{eqnarray}\label{F_omega_pi}
  F_{\omega\pi}(Q^2)
  &=&
  \left(
    e_uF_D^{(u)}(Q^2)+e_{\bar{d}}F_D^{(\bar{d})}(Q^2)
  \right)H_D^{\rho\pi}(Q^2)
  +\left(
     \kappa_uF_P^{(u)}(Q^2)
             +\kappa_{\bar{d}}F_P^{(\bar{d})}(Q^2)
   \right)H_P^{\rho\pi}(Q^2) \nonumber \\
  &=&
  F_D^{(q)}H_D^{\rho\pi}(Q^2)
  +\left(\kappa_u-\kappa_d\right)
   F_P^{(q)}(Q^2)H_P^{\rho\pi}(Q^2) 
,\end{eqnarray}
where the body form factors $H_D^{\rho\pi}$ and
$H_P^{\omega\pi}$ are given by~\cite{CGN95c}
\begin{equation}
  H_D^{\rho\pi}(Q^2)
  =2\int d^2\vec{k}_{\perp} \int dx
   \sqrt{\frac{\partial k_n}{\partial x}}
   \sqrt{\frac{\partial k_n'}{\partial x}}
   \phi_{1S}(x,\vec{k}_{\perp})
   \phi^*_{1S}(x,\vec{k}_{\perp}')
   \frac{m\lambda+2k_y^2}{xM_0M_0'\lambda}
\end{equation}
and
\begin{equation}
  H_P^{\rho\pi}(Q^2)=
   \int d^2\vec{k}_{\perp} \int dx
   \sqrt{\frac{\partial k_n}{\partial x}}
   \sqrt{\frac{\partial k_n'}{\partial x}}
   \phi_{1S}(x,\vec{k}_{\perp})
   \phi^*_{1S}(x,\vec{k}_{\perp}')
   \frac{\lambda
         \left(
           m^2+\vec{k}_{\perp}\vec{k}_{\perp}'
         \right) - 2M_0(1-x)k^2_y}
        {x(1-x)M_0M_0'm\lambda}        
.\end{equation}
Here, $\lambda=2m+M_0$.

We are now able to fix the parameters 
$\kappa_u$ and $\kappa_d$.
The value of the transition form factors at $Q^2=0$,
the transition magnetic moments
$\mu_{\rho\pi}$ and $\mu_{\omega\pi}$,
have been experimentally 
determined from the radiative decay widths of $\rho$ and
$\omega$ mesons~\cite{Jau91,ChJ97} viz. 
\begin{equation}
  \Gamma(\rho\to\pi\gamma)
  =
  \frac{1}{3}\alpha {F_{\rho\pi}(0)}^2
  \left(\frac{M_\rho^2-M_\pi^2}{2M_\rho}\right)^3
\end{equation}
and a similar expression for 
$\Gamma(\omega\to\pi\gamma)$
as $\mu_{\rho\pi}=0.741\pm0.038~\text{GeV}^{-1}$ and
$\mu_{\omega\pi}=2.33\pm0.06~\text{GeV}^{-1}$, 
respectively~\cite{Cas98}. 
From (\ref{F_rho_pi})
and (\ref{F_omega_pi}) we get
\begin{equation}
  \mu_{\rho\pi}
    =\frac{H_D^{\rho\pi}(0)}{3}
     +(\kappa_u+\kappa_d)H_P^{\rho\pi}(0)
\end{equation}
and
\begin{equation}
  \mu_{\omega\pi}
    =H_D^{\omega\pi}(0)
     +(\kappa_u-\kappa_d)H_P^{\omega\pi}(0)
.\end{equation}
By fitting the experimental value for 
$\mu_{\rho\pi}$ and $\mu_{\omega\pi}$ one 
gets
$\kappa_u-\kappa_d=0.138\pm0.015$ 
and 
$\kappa_u+\kappa_d=0.036\pm0.01$
or, correspondingly,
$\kappa_u=0.087\pm0.013$
and
$\kappa_d=-0.051\pm0.013$.

The mean square radii (MSR) 
associated with $F_D^{(q)}$ and $F_P^{(q)}$ are
$<{r_D^{(q)}}^2>=
  -6\left.\frac{dF_D^{(q)}}{dQ^2}\right|_{Q^2=0}$ 
and 
$<{r_P^{(q)}}^2>=
  -6\left.\frac{dF_P^{(q)}}{dQ^2}\right|_{Q^2=0}$. 
Assuming the same internal electromagnetic structure for
quark and antiquark, we have
$r_D^{(u)}=r_D^{(d)}\equiv r_D^{(q)}$ and 
$r_P^{(u)}=r_P^{(d)}\equiv r_P^{(q)}$.

To determine $r_D^{(q)}$ and $r_P^{(q)}$,
we fit
our model to the well known electromagnetic form factor for
the pion $F_\pi$, 
following the procedure outlined in~\cite{CGN95c}. 
Using (\ref{F_pi}), we get
\begin{equation} \label{E:MSR_pi}
  <r_{\pi}^2>=-6\left.\frac{dF_{\pi}}{dQ^2}\right|_{Q^2=0}
  =
  <r_D^2>+<r_{H_D^{\pi}}^2>
     +(\kappa_u-\kappa_d)<r_{H_P^{\pi}}^2>
,\end{equation}
where 
$<r_{H_D^{\pi}}^2>=
   -6\left.\frac{dH_D^{\pi}}{dQ^2}\right|_{Q^2=0}$ 
and
$<r_{H_P^{\pi}}^2>=
   -6\left.\frac{dH_P^{\pi}}{dQ^2}\right|_{Q^2=0}$ 
are the 
MSR associated with the body form factors $H_D^{\pi}$ and
$H_D^{\pi}$. Using our 
values for $\kappa_u$ and $\kappa_d$, we find
$<r_D^2>=0.0495\pm0.022~\text{fm}^{2}$ to obtain the 
experimental value
$<r_{\pi}^2>=0.432\pm0.016~\text{fm}^{2}$
~\cite{Ame86}.
The last parameter to be determined, $<r_P^2>$, is not
affected by the $Q^2\approx 0$ range of $F_{\pi}$ 
(see Eq.~(\ref{E:MSR_pi}))
and can be
chosen to be $<r_P^2>=0.136~\text{fm}^{2}$
to get a good agreement in the whole range of existing
pion form factor data.

Finally, making use of the uncertainties 
due to the experimental errors, we determine our
parametrization (see Table~\ref{T:parameters})
to get the best possible fit 
for the rho decay constant $f_{\rho}$.



\begin{references}

\bibitem{GoI85}
S. Godfrey and N. Isgur, Phys. Rev. {\bf D32},  189  (1985).

\bibitem{CGN94}
F. Cardarelli {\it et~al.}, Phys. Lett. {\bf B332},  1  (1994).

\bibitem{CGN95b}
F. Cardarelli {\it et~al.}, Few-Body Systems Suppl. {\bf 9},  267  (1995).

\bibitem{CGN95c}
F. Cardarelli {\it et~al.}, Phys. Lett. {\bf B359},  1  (1995).

\bibitem{CGN95d}
F. Cardarelli {\it et~al.}, Phys. Rev. {\bf D53},  6682  (1996).

\bibitem{CGN95a}
F. Cardarelli {\it et~al.}, Phys. Lett. {\bf B349},  393  (1995).

\bibitem{ChJ98}
H.-M. Choi and C.-R. Ji, Phys. Rev. {\bf D59},  074015  (1999).

\bibitem{BPP97}
S.~J. Brodsky, H.-C. Pauli, and S.~S. Pinsky, Phys. Rept. {\bf 301},  299
  (1998).

\bibitem{Dir49}
P.~M.~A. Dirac, Rev. Mod. Phys. {\bf 21},  392  (1949).

\bibitem{JiR96}
C.-R. Ji and S.-J. Rey, Phys. Rev. {\bf D53},  5815  (1996).

\bibitem{JiS92}
C.-R. Ji and Y. Surya, Phys. Rev. {\bf D46},  3565  (1992).

\bibitem{DrY70}
S.~D. Drell and T.-M. Yan, Phys. Rev. Lett. {\bf 24},  181  (1970).

\bibitem{Mel74}
H.~J. Melosh, Phys. Rev. {\bf D9},  1095  (1974).

\bibitem{ScI95}
D. Scora and N. Isgur, Phys. Rev. {\bf D52},  2783  (1995).

\bibitem{Cas98}
C. Caso {\it et~al.}, Eur. Phys. J. {\bf C3},  1  (1998).

\bibitem{Jau90}
W. Jaus, Phys. Rev. {\bf D41},  3394  (1990).

\bibitem{Jau91}
W. Jaus, Phys. Rev. {\bf D44},  2851  (1991).

\bibitem{Ger95}
S.~B. Gerasimov, Phys. Lett. {\bf B357},  666  (1995).

\bibitem{BBH78}
C.~J. Bebek {\it et~al.}, Phys. Rev. {\bf D17},  1693  (1978).

\bibitem{ChJ97}
H.-M. Choi and C.-R. Ji, Nucl. Phys. {\bf A618},  291  (1997).

\bibitem{BrH92}
S.~J. Brodsky and J.~R. Hiller, Phys. Rev. {\bf D46},  2141  (1992).

\bibitem{BaE85}
A.~S. Bagdasaryan and S.~V. Esaybegyan, Sov. J. Nucl. Phys. {\bf 42},  278
  (1985).

\bibitem{BeJ69}
J.~S. Bell and R. Jackiw, Nuovo Cimento {\bf A60},  47  (1969).

\bibitem{Adl69}
S.~L. Adler, Phys. Rev. {\bf 177},  2426  (1969).

\bibitem{BJP98}
S.~J. Brodsky, C.-R. Ji, A. Pang, and D.~G. Robertson, Phys. Rev. {\bf D57},
  245  (1998).

\bibitem{CCP88}
P.~L. Chung, F. Coester, and W.~N. Polyzou, Phys. Lett. {\bf B205},  545
  (1988).

\bibitem{Dzi88}
Z. Dziembowski, Phys. Rev. {\bf D37},  778  (1988).

\bibitem{Ame86}
S.~R. Amendolia {\it et~al.}, Phys. Lett. {\bf B178},  435  (1986).

\bibitem{Cel91}
{CELLO coll., H.-J. Behrend} {\it et~al.}, Z. Phys {\bf C49},  401  (1991).

\bibitem{Cel95}
{CELLO coll., A. Savinov} {\it et~al.}, PHOTON 95 Conference, Sheffield
  (1995).


\end{references}

%
%

%
%

\widetext
\begin{table}
  \caption{Model parameters for the best fit of the mass spectrum
           for both HO and
           linear potential for several quark masses $m_q$.  
           The parameter $\beta$ has been obtained 
           from the variational principle for the ground state.
           For comparison, the parametrization 
           from~\protect\cite{ChJ98}
           together with the ground state masses are shown.
           Also
           shown is the pion decay constant $f_{\pi}$ used to
           distinguish between the different parametrizations.
           All units are in GeV, except the parameter $b$ which is in
           $\text{GeV}^3$ [$\text{GeV}^2$] for HO [linear] potential,
           $f_{\pi}$ which is in MeV and $\kappa$ which is 
           dimensionless.}
  \label{T:mass_spectrum}
  \begin{tabular}{c | c  c  c  c  c  c  c  c  c  c  c}
      & HO & HO & HO & HO & HO & HO & linear & linear 
      & HO\tablenotemark[1] & linear\tablenotemark[1] 
      & Experiment\tablenotemark[2] \\
    \hline
    $m_q$
      & 0.150 & 0.180 & 0.190 & 0.200 & 0.250 & 0.300 
      & 0.150 & 0.250 & 0.250 & 0.220 & \\
    $a$ 
      & 0.268 & 0.264 & 0.256 & 0.246 & 0.243 & 0.220 
      & 0.112 & 0.187 & -0.144 & -0.724 & \\
    $b$ 
      & 0.0127 & 0.0117 & 0.0118 & 0.012  & 0.0092 & 0.0069 
      & 0.089 & 0.05 & 0.010 & 0.18 & \\
    $\kappa$ 
      & 1.207 & 1.22 & 1.22 & 1.218 & 1.246 & 1.26 
      & 1.165 & 1.242 & 0.607 & 0.313 & \\
    \hline
    $\beta$ 
      & 0.4782 & 0.4894 & 0.4957 & 0.5019 & 0.5224 & 0.5463 
      & 0.4914 & 0.5277 & 0.3194 & 0.3659 & \\
    $M_\pi$ 
      & 0.136 & 0.139 & 0.138 & 0.138 & 0.138 & 0.137 & 0.139 
      & 0.138 & 0.137 & 0.135 & 0.135$\ldots$0.139 \\
    $M_{\pi(2S)}$ 
      & 1.280 & 1.272 & 1.275 & 1.279 & 1.261 & 1.253 
      & 1.288 & 1.258 & & & 1.3$\pm$0.1 \\
    $M_{\pi(3S)}$ 
      & 1.799 & 1.797 & 1.806 & 1.817 & 1.802 & 1.805 
      & 1.781 & 1.776 & & & 1.801$\pm$0.013 \\
    $M_{\rho}$ 
      & 0.770 & 0.770 & 0.0770 & 0.770 & 0.770 & 0.770 
      & 0.770 & 0.770 & 0.770 & 0.770 & 0.770$\pm$0.0008 \\
    $M_{\rho(2S)}$ 
      & 1.437 & 1.441 & 1.448& 1.456 & 1.459 & 1.470 
      & 1.442 & 1.457 & & & 1.465$\pm$0.025 \\
    $M_{\rho(3S)}$ 
      & 2.051 & 2.049 & 2.059 & 2.071 & 2.061 & 2.069 
      & 2.033 & 2.037 & & & \tablenotemark[3] \\
    \hline
    $f_{\pi}$
      & 75.8 & 87.9 & 91.9 & 95.8 & 113.9 & 129.9 & 76.2 
      & 113.9 & 92.4 & 91.8 & 92.4$\pm$0.25
  \end{tabular}
  \tablenotetext[1]{Predictions from~\cite{ChJ98}.}
  \tablenotetext[2]{Data taken from~\cite{Cas98}.}
  \tablenotetext[3]{See the text for a discussion 
                    of possible resonances at
                    $1.700\pm0.020~\text{GeV}$ and
                    $2.149\pm0.017~\text{GeV}$.}
\end{table}

\narrowtext
\begin{table}
  \caption{The model parameters.}
  \label{T:parameters}
  \begin{tabular}{c c}
    $m_q=0.190~\text{GeV}$ & $<r_D^2>=0.050~\text{fm}^2$ \\
    $a=0.256~\text{GeV}$ & $ <r_P^2>=0.136~\text{fm}^2$ \\
    $b=0.0118~\text{GeV}^3$ & $\kappa_u=0.074$ \\
    $\kappa=1.22$ & $\kappa_d=-0.048$ \\
    $\beta=0.4957~\text{GeV}$ & \\
  \end{tabular}
\end{table}

\mediumtext
\begin{table} 
  \caption{Observables for different model parameters.
           The first column shows the calculations
           from~\protect\cite{ChJ97,ChJ98}.
           The second and third columns show our
           predictions without and with CQ form factors. }
  \label{T:observables}
  \begin{tabular}{c | c  c  c  c}
    & Choi/Ji~\cite{ChJ97,ChJ98} & our w/o CQ FF
       & our w/ CQ FF & Experiment \\
    \hline
    $m$ [GeV] & 0.250 & 0.190 & 0.190 & \\
    $\beta$ [GeV] & 0.3194 & 0.4957 & 0.4957 & \\
    $<r_D^2> [\text{fm}^2]$ & 0 & 0 & 0.050 & \\
    $<r_P^2> [\text{fm}^2]$ & 0 & 0 & 0.136  & \\
    $\kappa_u$ & 0 & 0 & 0.074 &\\
    $\kappa_d$ & 0 & 0 & -0.048 & \\
    \hline
    $f_\pi$[MeV] & 92.4 & 91.9 & 91.9 & 92.4$\pm$0.025 \\
    $<r_\pi^2> [\text{fm}^2]$ & 0.448 & 0.327 &
             0.427 & 0.432$\pm$0.016  \\
    $\Gamma_{Anom}(\pi\to\gamma^*\gamma)$ [eV] & 7.73 & 7.82 & 7.82 & 
             7.25$\pm$0.23 \\
    $f_\rho$ [MeV] & 151.9 & 231.7 & 153.0 & 152.9$\pm$3.6 \\
    $\mu_{\pi^{+}\rho^{+}} [\text{GeV}^{-1}]$ & 0.782 & 0.603 & 0.701 &
             0.741$\pm$0.038 \\
    $\mu_{\pi^{0}\omega} [\text{GeV}^{-1}]$ & 2.35 & 1.81 & 2.27 &
             2.33$\pm$0.06 \\
    $\rho:\mu_1$ & 2.2 & 2.27 & 2.63 & \tablenotemark[1] \\
    $\rho: Q_1$ & 0.20 & 0.53 & 0.69 & \tablenotemark[1]
    \end{tabular}
    \tablenotetext[1]{
      There are no data available yet. However, the  results
      of other theoretical calculations are given by:
      $\mu_1:~2.26,~2.1,~2.3$ and 
      $Q_1:~0.37,~0.41,~0.45$
      from~\cite{CGN95a}, \cite{ChJ97} and 
      \cite{BaE85}, respectively.
    }
\end{table}

\narrowtext
\begin{table}
  \caption{Shows predictions of several hadronic
           properties for radially excited states.
           $\mu_{{\rho'}^+\pi^+}$ and $\mu_{\omega'\pi^0}$
           are transition magnetic moments for the 
           transition from the first excited state ($2S$)
           to the ground state ($1S$).}
  \label{T:predictions}
  \begin{tabular}{c | c c c}
    & $1S$ & $2S$ & $3S$ \\
    \hline
    $f_\pi$[MeV] & 91.9 & 18.6 & 42.3 \\
    $<r_\pi^2> [\text{fm}^2]$ & 0.427 & 0.606 & 0.854 \\
    $f_\rho$ [MeV] & 153.0 & 90.5 & 96.4 \\
    $\mu_{\pi^{+}\rho^{+}} [\text{GeV}^{-1}]$ & 0.70 & 0.55 & 0.48 \\
    $\mu_{\pi^{0}\omega} [\text{GeV}^{-1}]$ & 2.27 & 1.79 & 1.58 \\
    \hline
    $\mu_{{\rho'}^+\pi^+} [\text{GeV}^{-1}]$ & 
      \multicolumn{2}{c}{-0.23} &  \\
    $\mu_{\omega'\pi^0} [\text{GeV}^{-1}]$ & 
      \multicolumn{2}{c}{-0.71} &  \\
  \end{tabular}
\end{table}

\end{document}